\documentclass[aip,reprint,amsmath,amssymb]{revtex4-1}
\usepackage{graphicx}
\usepackage{caption}
\usepackage{subcaption}

\begin{document}

\title{Morphology of Shocked Lateral Outflows in Colliding Hydrodynamic Flows} 

\author{R.N. Markwick}
\author{A. Frank}
\author{J. Carroll-Nellenback}
\author{E.G. Blackman}
\affiliation{Department of Physics and Astronomy, University of Rochester, Rochester, NY 14627}
\author{P.M. Hartigan}
\affiliation{Department of Physics and Astronomy, Rice University, Houston,  TX, 77005-1892}
\author{S.V. Lebedev}
\author{D.R. Russell}
\author{J.W.D. Halliday}
\author{L.G. Suttle}
\affiliation{Blackett Laboratory, Imperial College, London SW7 2BW, United Kingdom}

\begin{abstract}
Supersonic interacting flows occurring in phenomena such as protostellar jets give rise to strong shocks, and have been demonstrated in several laboratory experiments.
To study such colliding flows, we use the AstroBEAR AMR code to conduct hydrodynamic simulations in three dimensions.
We introduce variations in the flow parameters of density, velocity, and cross sectional radius of the colliding flows
in order to study the propagation and conical shape of the bow shock formed by collisions between two, not necessarily symmetric, hypersonic flows. 
We find that the motion of the interaction region is driven by imbalances in ram pressure between the two flows, while the conical structure of the bow shock is a result of shocked lateral outflows (SLOs) being deflected from the horizontal when the flows are of differing cross-section.
\end{abstract}


\maketitle 

\section{Introduction}

Radiative shocks occur in a variety of settings, such as High Energy Density Plasma (HEDP), protostellar jets \citep{protostellarJet}, supernova explosions \citep{Supernova}, gamma ray bursts \citep{GRB}, and active galactic nuclei \citep{AGN}. 
 Over the years several efforts have been made to study these phenomena in the laboratory. Despite large differences in physical scale, results obtained in the laboratory can be used to understand phenomena in the astrophysical setting, thanks to the use of dimensionless parameters \citep{ryutov2000, ryutov2001, Falize2011}.
Experiments at Imperial College London have produced single flows using a pulsed-power generator to drive radial wire arrays \citep{Lebedev05} and  radial foils \citep{suzukiVidal09, Ciardi}; similar experiments have also been performed at Cornell University \citep{gourdain10}.
Collimation of these flows was driven by toroidal magnetic fields in a manner consistent with the ``Magnetic Tower" model \citep{LyndenBell96}. Extending this work, \citet{suzukiVidal12} found that the presence of an ambient medium results in the formation of a shock, driven by ablation of halo plasma.

\begin{figure}
    \centering
        \includegraphics[width=0.68\columnwidth]{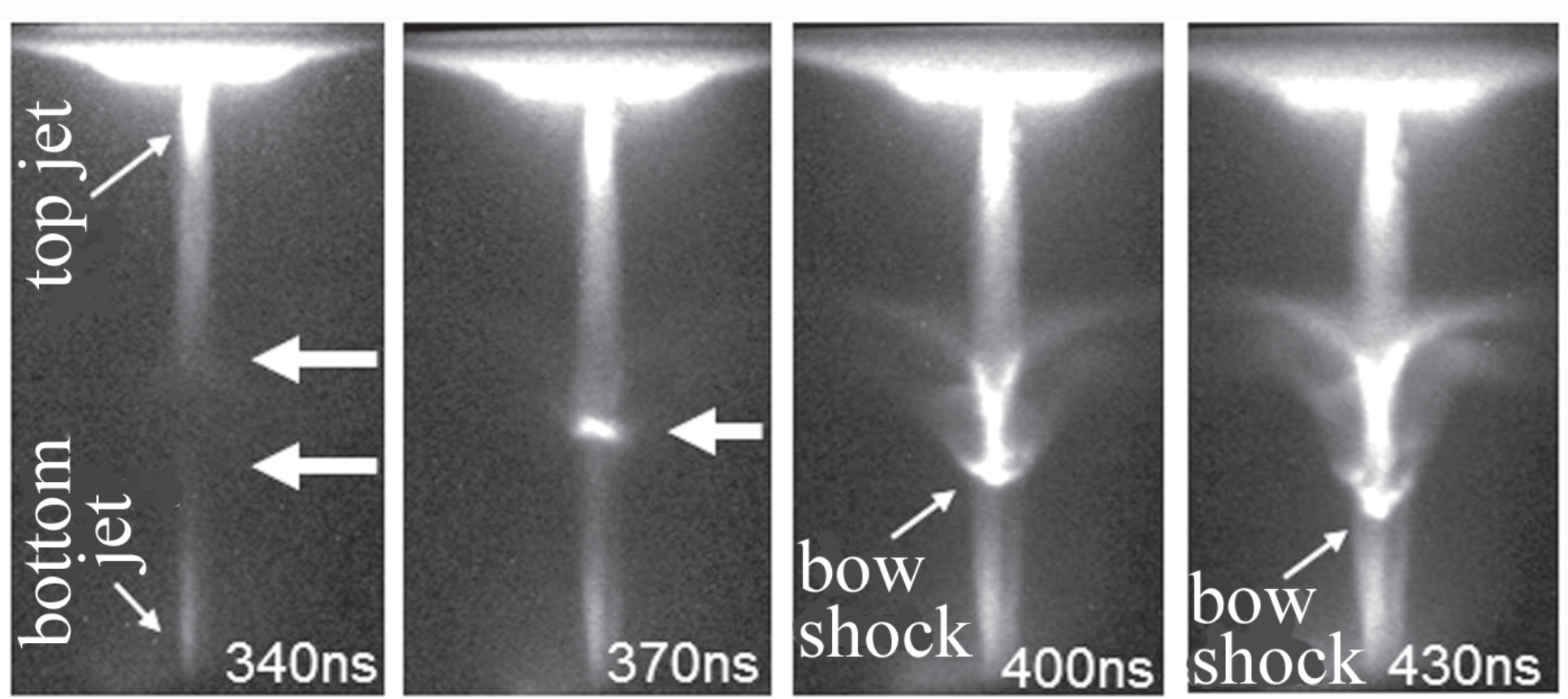}
        \includegraphics[width=0.29\columnwidth]{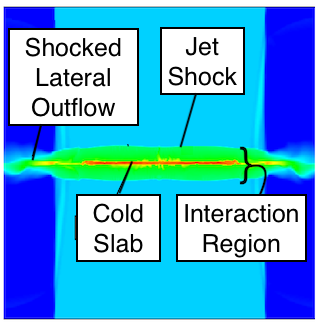}
    \caption{(left) Images of a colliding flow experiment. Adapted from figure 2 of  \citet{suzukiVidal15}.\\
    (right) a diagram showing the structure of colliding flows; adapted from \citet{Markwick21} }
    \label{fig:SV}
\end{figure}

While astronomical shocks are generally produced via time-variations in velocity, a similar effect can be obtained in a laboratory setting by creating a collision between two supersonic flows.
 \citet{suzukiVidal15} conducted such an experiment, images of which are shown in figure \ref{fig:SV}. 
Among the results of that study are the emergence of small scale structures behind a conical bow shock, which was observed to propagate in a downward direction.

When two flows collide, an \textit{interaction region} forms, with the jet shocks becoming the boundaries. This region usually consists of a \textit{cooling region}, in which gas cools from post-shock temperature $T_s$ immediately behind the jet shocks to a lower temperature further behind the shock. 
 After passing through the cooling region, gas reaches its final post-cooling temperature and is deposited onto the \textit{cold slab}, where densities are highest (see figure \ref{fig:SV}), some shocked material gets ejected laterally by the high pressures throughout the interaction region, producing shocked lateral outflows (SLOs) \citep{Raga93}. 

In our previous paper \citep{Markwick21} we began a series of simulations in an effort to study colliding radiative flows like those of \citet{suzukiVidal15}. In that paper we used hydrodynamic simulations with an analytic form of radiative cooling, focusing our attention on instabilities in an effort to explain the origin of small scale structures. While \citet{suzukiVidal15} had suggested that the \citet{field65} instability was a source of structure, we did not see any evidence of this mode present in the case of analytic cooling. Instead, long-term evolution was found to be dominated by the bending modes characteristic of the nonlinear thin shell instability (NTSI) \citep{Vishniac94}, which could be triggered either by sufficiently short cooling lengths or by oscillations resulting from the radiative shock instability \citep{Langer81}.

In this paper, we continue to investigate the results of \citet{suzukiVidal15} by extending the results of \citet{Markwick21}. Here we shift our focus away from instabilities (which were discussed in significant depth in \citet{Markwick21}) to a study of flow parameter variation, allowing us to better understand the conical shape and downward propagation of the bow shock and lateral outflows.
We will continue to focus solely on hydrodynamic simulations, but the analytic cooling curve previously used will be replaced with a more realistic cooling function; this will be described in further detail in section \ref{sec:meth}. Our approach of continuing to remain in the radiative hydrodynamic case for now will eventually allow us to have a better understanding of the magnetic case by providing a reference for comparison.

This paper is organized as follows: 
In Section \ref{sec:meth} we discuss the model system, the selected cooling function, and simulation parameters.
In section \ref{sec:result}, we present the results of the simulations. Section \ref{sec:discuss} will include a discussion of motion of the interaction region, properties of the cold slab, and deflection of the SLOs.

\section{Methods and Model} \label{sec:meth}

The simulations in this study were conducted using AstroBEAR\footnote{https://astrobear.pas.rochester.edu/}${ }^{,}$ \citep{cunningham09,carroll13}, which is a massively parallelized adaptive mesh refinement (AMR) code that includes a variety of multiphysics solvers, such as magnetic resistivity, radiative transport, self-gravity, heat conduction, and ionization dynamics.  Our study uses only the hydro solvers with an energy source term associated with the radiative cooling. Thus our governing equations are;
\begin{equation}
    \frac{\partial \rho}{\partial t} + \boldsymbol{\nabla} \cdot \rho \boldsymbol{v} = 0 
    \label{eq:Eu1}
\end{equation}
\begin{equation}
    \frac{\partial \rho \boldsymbol{v}}{\partial t} + \boldsymbol{\nabla} \cdot \left ( \rho \boldsymbol{v} \otimes \boldsymbol{v} \right )= - \nabla p
    \label{eq:Eu2}
\end{equation}
\begin{equation} 
    \frac{\partial E}{\partial t} + \boldsymbol{\nabla} \cdot ((E + p) \boldsymbol{v}) =-n^2\Lambda(n,T)
    \label{eq:Eu3}
\end{equation}
where $\rho$ is the mass density, $n$ is the number density of nuclei, $\boldsymbol{v}$ is the fluid velocity, $p$ is the thermal pressure,  and $E = \frac{1}{\gamma - 1} p + \frac{1}{2}\rho v^2$ is the combined internal and kinetic energies. $n^2\Lambda(n,T)$ is a cooling function, which gives the energy loss per unit volume.

\begin{figure}
	\includegraphics[width=\columnwidth]{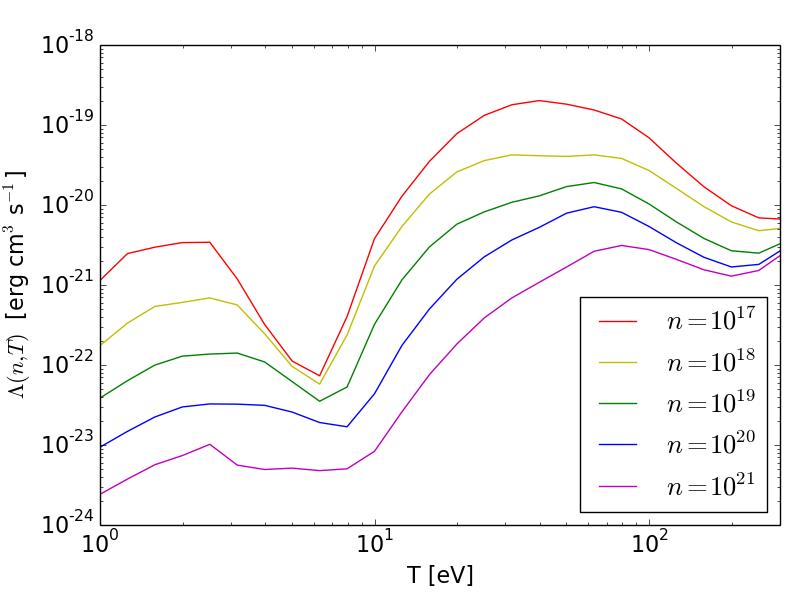}
    \includegraphics[width=\columnwidth]{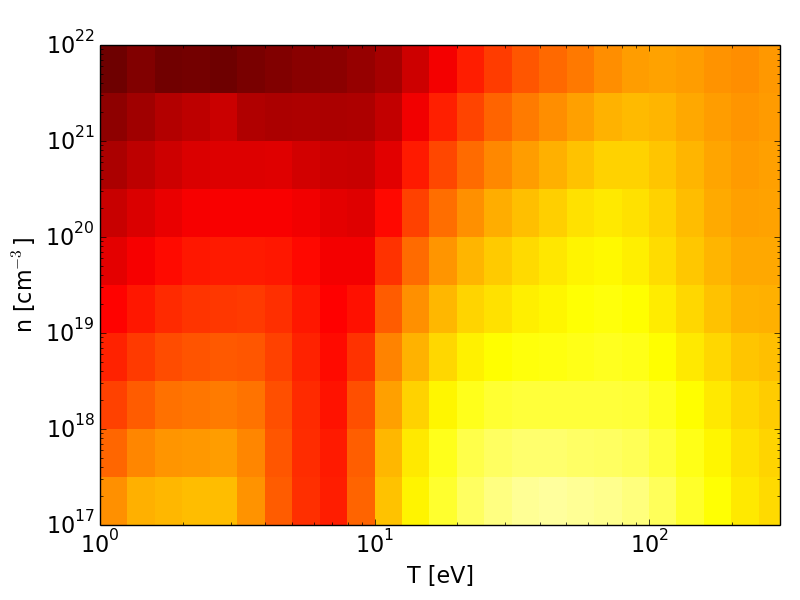}
    \caption{
    The cooling function plotted (top) as a function of temperature, for selected densities, and (bottom) as a colour-map in terms of density and temperature.
    }
    \label{fig:cool}
\end{figure}

While in \citep{Markwick21} we used an analytic cooling function of the form $\Lambda(T) = \alpha \left(\frac{T}{T_0}\right)^\beta$, in this paper we use a more realistic cooling function.  Such a cooling function introduces variation in the slope $\frac{\operatorname{d} \log\Lambda}{\operatorname{d}\log T}$, which is important to both the radiative shock instability and to the Field instability.
The specific function $\Lambda(n,T)$ we used was chosen in attempt to provide correspondence with the experiments of \citet{suzukiVidal15} and is plotted in figure \ref{fig:cool}.
This function was calculated using ABAKO/RAPCAL  \citep{RAPCAL08,RAPCAL09}.  The required atomic data were obtained using  FAC \citep{FAC}, while the resulting function was parameterized in terms of density and temperature using PARPRA \citep{PARPRA}.
Cooling is strongest at high temperatures and lower densities; as a fluid cools, $\Lambda(n,T)$ decreases until a local minimum is encountered at a temperature on the order of 10 eV. Cooling then increases again as temperature continues to decrease, but this effect is reduced at higher densities.

Similarly to \citet{Markwick21}, we drove two cylindrical jets, one from the top and bottom z-boundaries respectively. For the first run, both jets had densities of $3.0\times 10^{18}$ particles per cm$^3$, speeds of 70.0 km s$^{-1}$, temperature of 720 K, and cross sectional radii of 1.5 mm.  
The ambient medium  density was $1\times 10^{18}$ particles per cm$^3$ at a temperature of 4320 K. 
Extrapolated boundary conditions were used in all directions.

Runs were conducted in a 6.4mm cubic space, and consisted of the reference run described above and nine other runs in which jet parameters were varied as follows (summarised in table \ref{tab:params}).
For the first two runs, we changed the density of the bottom jet to $2.0\times 10^{18}$ and $1.0\times 10^{18}$ particles per cm$^3$.
For third and fourth runs, the radius of the bottom jet was varied to 2.0 and 1.0 mm, and the density was varied to $1.69\times 10^{18}$ and  $6.75\times 10^{18}$ particles per cm$^3$ such that the mass of the jets remained unaltered. 
For the fifth and sixth runs, the radius of the bottom jet was varied to 2.0 and 1.0 mm, but this time the density of the jet was reverted to the original value.
Finally, the seventh and eighth runs varied velocities of the top and bottom jets to 80 and 60 km s$^{-1}$ respectively; these values were chosen to match the experiments in \citet{suzukiVidal15}.  The seventh run retained the densities of the reference run, while the eighth run used densities of $2.63\times 10^{18}$ and $3.50\times 10^{18}$ so that the jets were of equal momentum density, and the ninth run used densities of $2.30\times 10^{18}$ and $4.08\times 10^{18}$ so that the jets were of equal momentum flux.

\begin{table}
    \centering
    \begin{tabular}{|c||c|c|c|c|}\hline
        Run  & $n_\text{jet}$ ($10^{18}$ cm$^{-3}$) & $v_\text{jet}$ (km s$^{-1}$) & $r_\text{jet}$ (mm) & Figures\\\hline\hline
         0  & 3.00/3.00 & 70/70 & 1.5 / 1.5& \ref{fig:1}-\ref{fig:4} \\\hline
         1& 3.00/2.00 & 70/70 & 1.5 / 1.5  & \ref{fig:2} \\\hline
         2 & 3.00/1.00 & 70/70 & 1.5 / 1.5& \ref{fig:2} \\\hline
         3& 3.00/1.69 & 70/70 & 1.5 / 2.0  & \ref{fig:3}\\\hline
         4 & 3.00/6.75 & 70/70 & 1.5 / 1.0 & \ref{fig:3}\\\hline       
         5 & 3.00/3.00 & 70/70 & 1.5 / 2.0 & \ref{fig:4}\\\hline
         6 & 3.00/3.00 & 70/70 & 1.5 / 1.0 & \ref{fig:4}\\\hline
         7 & 3.00/3.00 & 80/60 & 1.5 / 1.5 & \ref{fig:5}\\\hline
         8 & 2.63/3.50 & 80/60 & 1.5 / 1.5 & \ref{fig:5}\\\hline
         9 & 2.30/4.08 & 80/60 & 1.5 / 1.5 & \ref{fig:5},\ref{fig:6}\\\hline
    \end{tabular}
    \caption{Densities, velocities, and radii of jets in the first set of runs. For each pair of numbers, the first describes the top jet while the second describes the bottom jet.}
    \label{tab:params}
\end{table}

\section{Results} \label{sec:result}

\begin{figure}
	\includegraphics[width=\columnwidth]{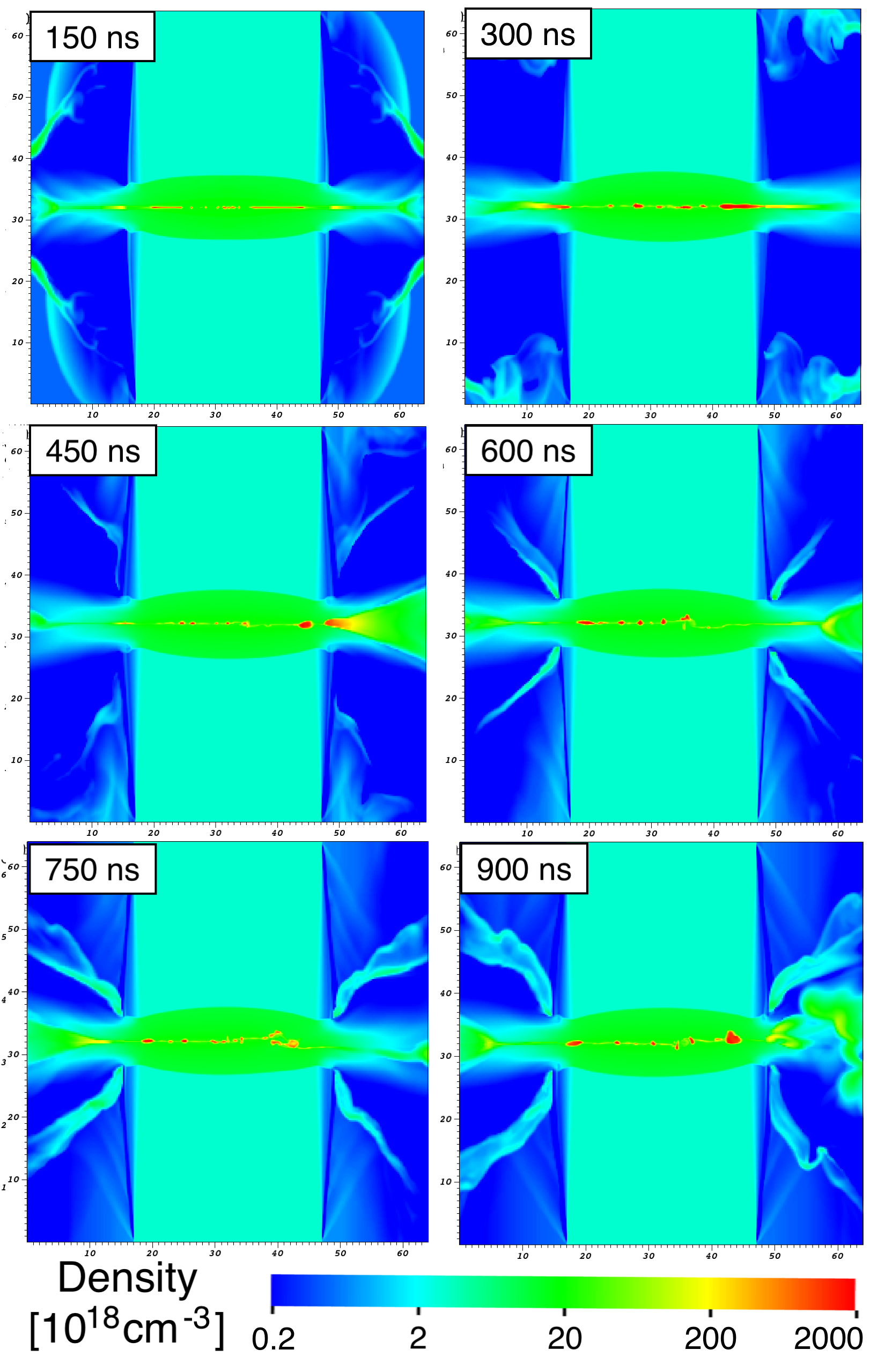}
    \caption{
    This and all subsequent figures (except as noted) show logarithmic density  midplane slices, with the major tick marks placed at intervals of 1mm. For this figure, the reference run  (run 0 in table \ref{tab:params}) is shown.}
    \label{fig:1}
\end{figure}

We will first examine the reference run (run 0 in table \ref{tab:params}), shown in figure \ref{fig:1}.  This run shows a collision between two identical jets, similar to those studied in \citet{Markwick21} but now using  the complex cooling curve described in section \ref{sec:meth}.
As gas crosses the jet shock and enters the interaction region, the density increases from $3\times 10^{18}$ cm$^{-3}$ to $1.2\times 10^{19}$ cm$^{-3}$. The density varies as the gas cools but remains at the same order of magnitude throughout the cooling region, but is seen to be as high as $2.5\times 10^{21}$ cm$^{-3}$ in the cold slab. As gas exits the interaction region through the sides and into the shocked lateral outflows, densities drop to a value of order $1\times 10^{18}$ cm$^{-3}$ and lower as gas begins to mix into the ambient medium. 
At later times clumps are observed to form in the cold slab, though the origin of these is uncertain given that the cold slab is not well resolved at just  a few cells ($10^{-3}$ cm) in width.
Since the jets are identical, the interaction region remains stationary and shows two identical jet shocks; while the shocks exhibit some curvature,  no conical structure is observed. The shocked lateral outflows flow primarily horizontally even though some material leaks from the top and bottom of the SLOs.

\begin{figure}
	\includegraphics[width=\columnwidth]{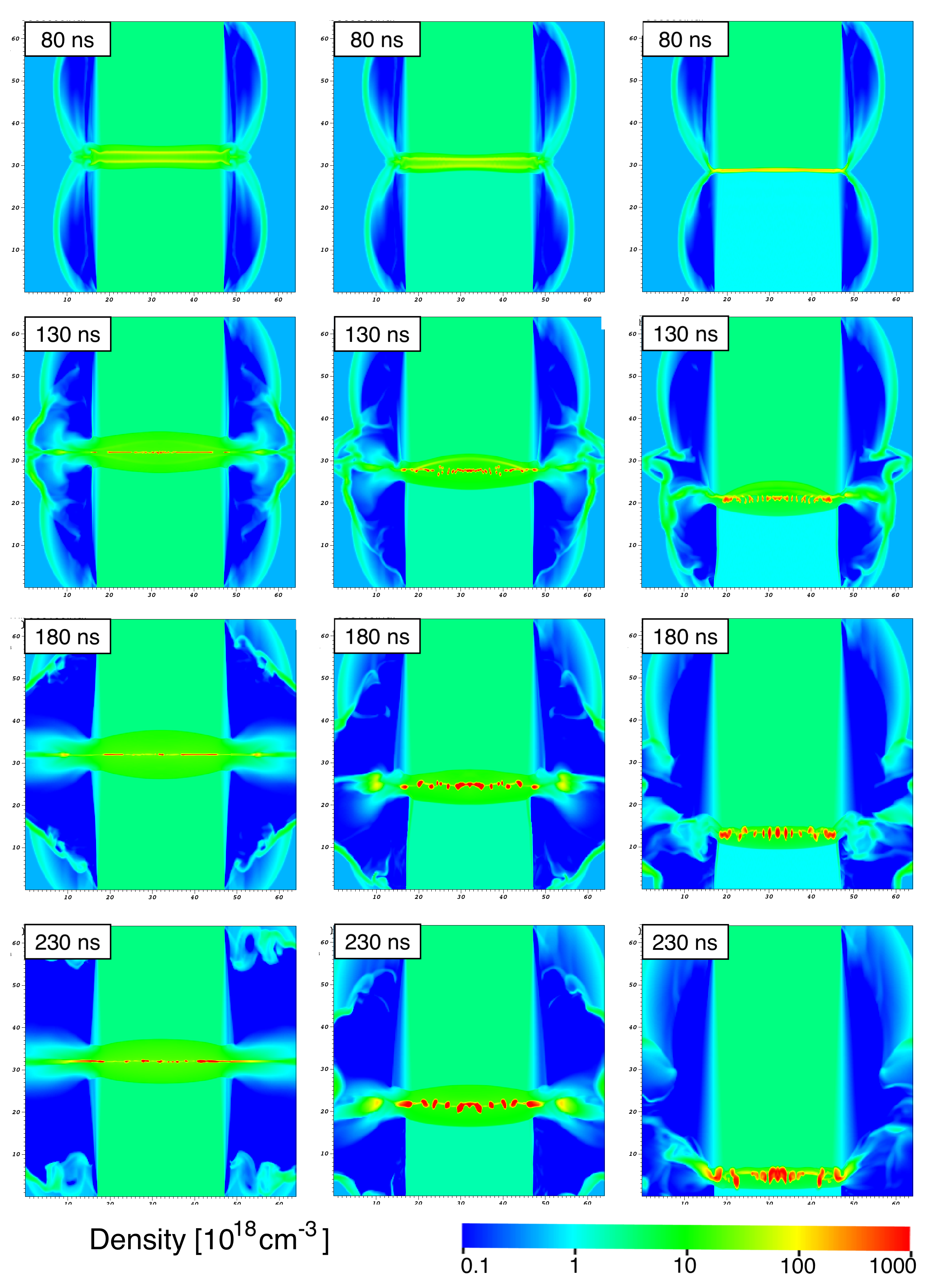}
    \caption{
    From left to right, runs 0, 1, and 2 from table \ref{tab:params}. For these runs, the density of the top jet remains fixed while the density of the bottom jet decreases with each successive run.
    }
    \label{fig:2}
\end{figure}

Next, we examine cases in which the density of the bottom jet is varied, while the density of the top jet is fixed at $1.0\times 10^{18}$ particles per cm$^3$. Figure \ref{fig:2} shows jets with densities $3.0\times 10^{18}, 2.0\times 10^{18}$, and $1.0\times 10^{18}$ particles per cm$^3$.  
As the density of the jets becomes imbalanced, the interaction region no longer remains in place, with the velocity of the interaction region in this case being proportional to the density imbalance. Lateral outflows are seen to be nearly horizontal. 
Clump formation in the cold slab begins much sooner in collisions with a density imbalance; while the details of smaller structures are beyond the scope of this paper, we conjecture that this is the result of asymmetry exacerbating the formation of perturbations within the cold slab.

Increasing the imbalance further does not significantly change the onset of clump formation. 
These clumps merge with each other as time progresses, though the regions of cold slab between these clumps is once again not well resolved.
Study of the the long term behaviour of these clumps is also limited by the interaction region hitting the wall after a relatively short amount of time.

\begin{figure}
	\includegraphics[width=\columnwidth]{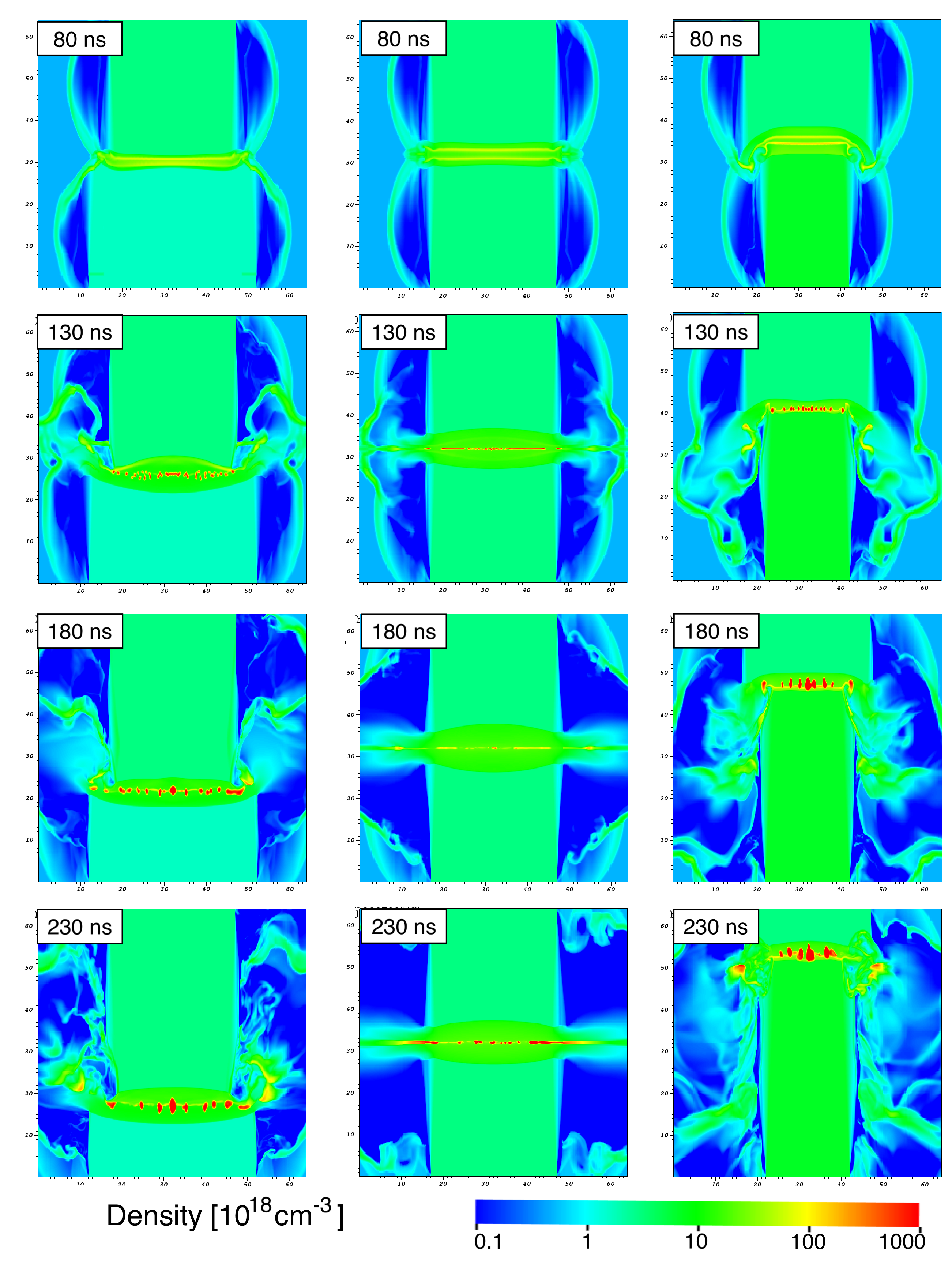}
    \caption{
     From left to right, runs 3, 0, and 4 from table \ref{tab:params}. For these runs, the density and radius of the top jet remains fixed while the bottom jet both increases in density and decreases in radius with each successive run.
    }
    \label{fig:3}
\end{figure}

Next, we examine cases in which the density of the bottom jet is again varied, this time varying the radius of the bottom jet as $r \propto \rho^{-\frac12}$ so that the total mass flux through a horizontal slice remains fixed. Figure \ref{fig:3} shows jets with densities $1.69\times 10^{18}, 3.0\times 10^{18}$, and $6.75\times 10^{18}$ particles per cm$^3$ and radii 2.0, 1.5, and 1.0 mm. The jet with the smaller radius, having a higher density, drives the motion of the interaction region. As with the previous set, an imbalance of density also results in formation of clumps within the cold slab at earlier times, with these clumps again merging into fewer but larger clumps as time progresses. 
Since the jets are of differing radii, the SLOs are no longer horizontal, instead bending away from the jet of larger radius.

\begin{figure}
	\includegraphics[width=\columnwidth]{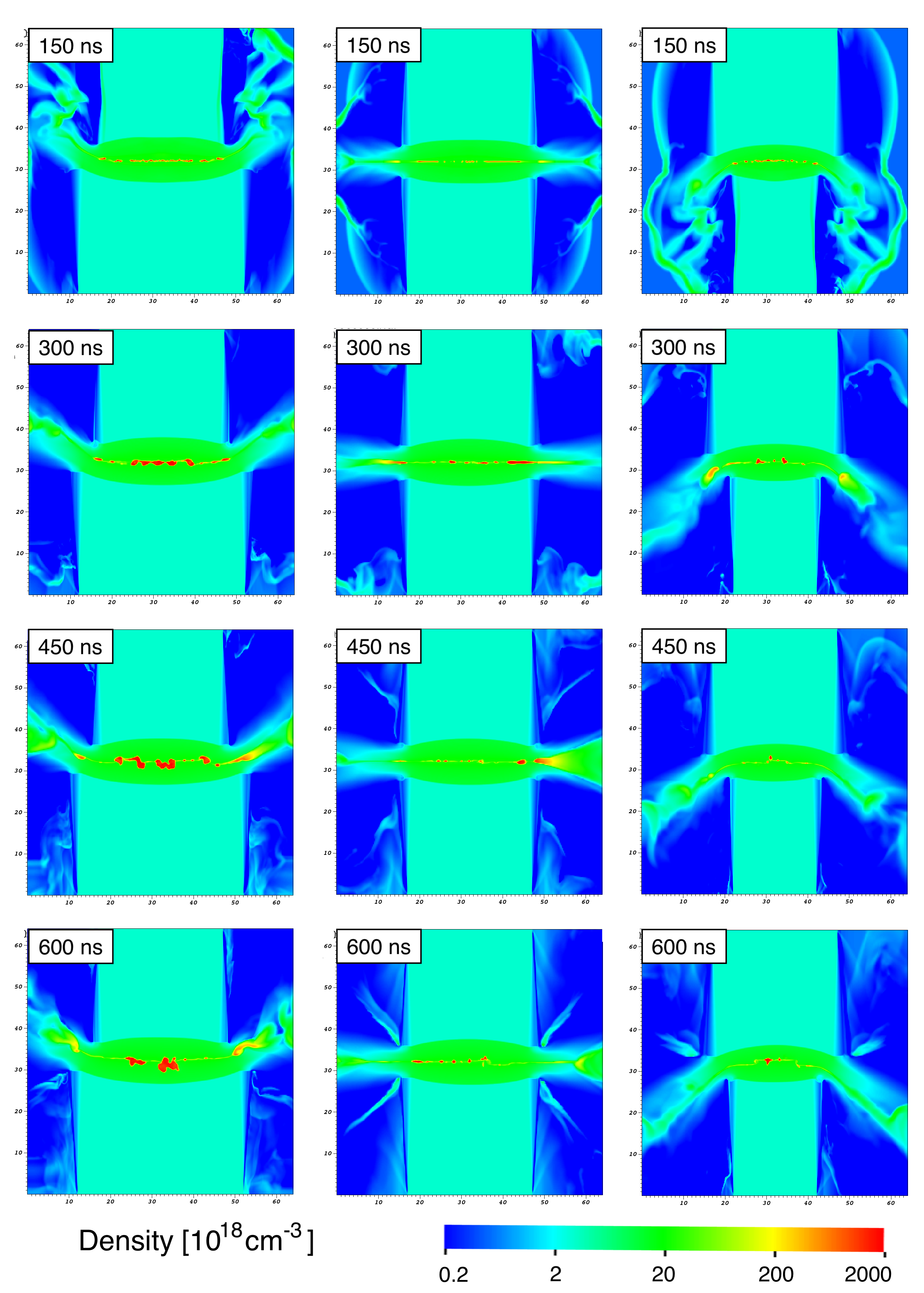}
    \caption{
    From left to right, runs 5, 0, and 6 from table \ref{tab:params}. For these runs, the radius of the top jet remains fixed while the radius of the bottom jet decreases with each successive run.
    }
    \label{fig:4}
\end{figure}

We next examine cases of equal speed jets for which the radius of the bottom jet is varied at fixed density.
Figure \ref{fig:4} shows jets with radii 2.0, 1.5, and 1.0 mm. For equal velocities and equal opposing jet densities, ram pressure is balanced and
the interaction region remains at a fixed location. 
As with density imbalances, an imbalance of radius alone appears to be sufficient to promote earlier clump formation, though in the case with a 1.5 mm jet colliding with a 1.0 mm jet the merging of clumps appears to be inhibited.

Since the speeds and
densities are balanced and the interaction region never approaches a wall, these cases can be observed for longer times. After material ejected in the initial collision flies away, a lateral outflow remains. When the jets differ in size, the direction of the SLO occurs at an angle to the horizontal plane and points away from the larger jet. The exact angle increases with the ratio $\frac{(r_+^2 - r_-^2)}{r_-}$, where $r_+$ and $r_-$ are the radii of the larger and smaller jet respectively, and is explored in section \ref{sec:angle}. The interaction region also exhibits curvature, bending away from the larger flow.

\begin{figure}
	\includegraphics[width=\columnwidth]{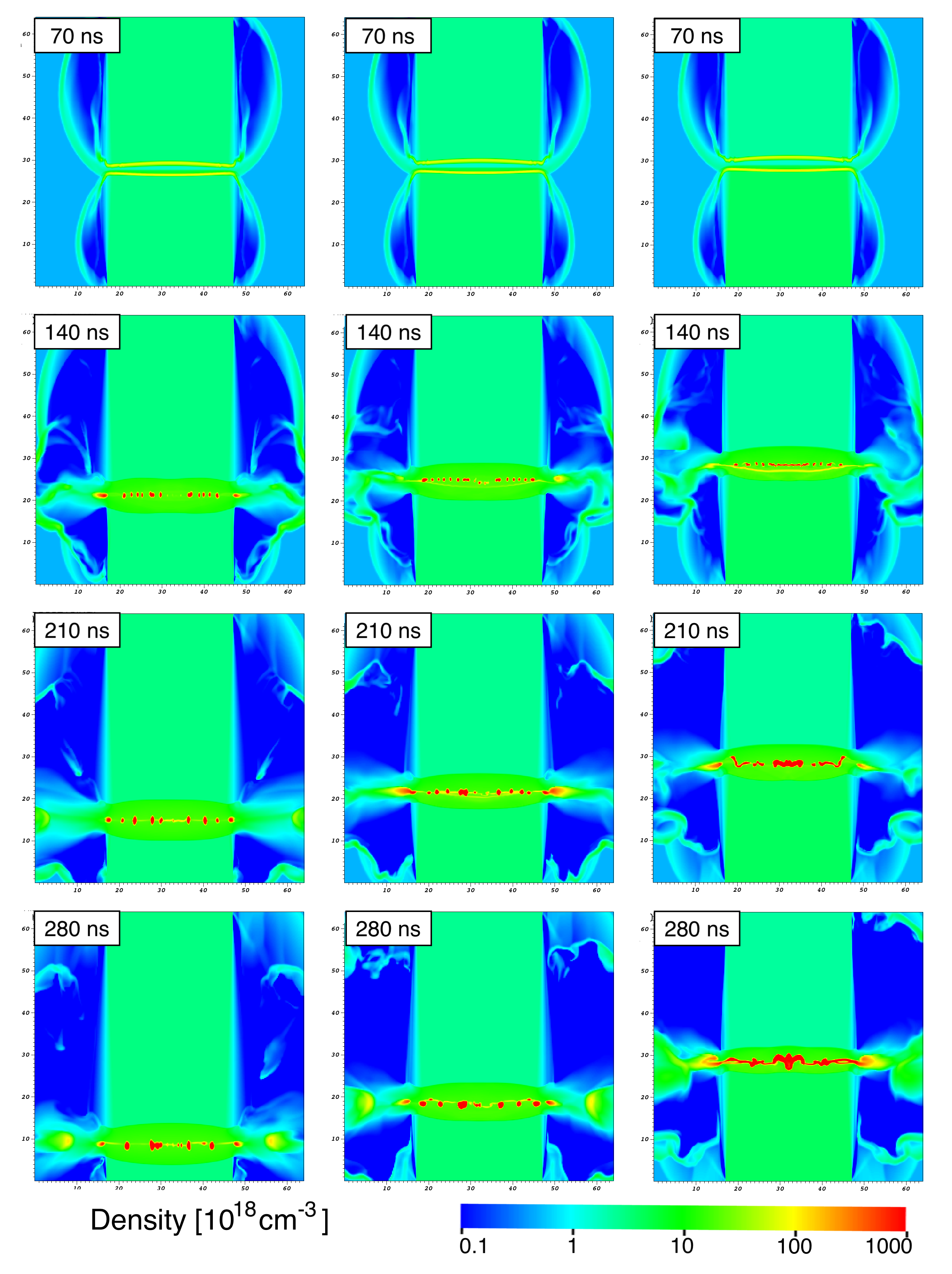}
    \caption{
    From left to right, runs 7, 8, 9 from table \ref{tab:params}. While the previous runs use $v = 70$ km s$^{-1}$ for both jets, these runs use jet velocities of 80 and 60 km s$^{-1}$  for the top and bottom jets respectively. The density of the top jet decreases (and the bottom jet increases) with  each successive run.
    }
    \label{fig:5}
\end{figure}

\begin{figure}
	\includegraphics[width=\columnwidth]{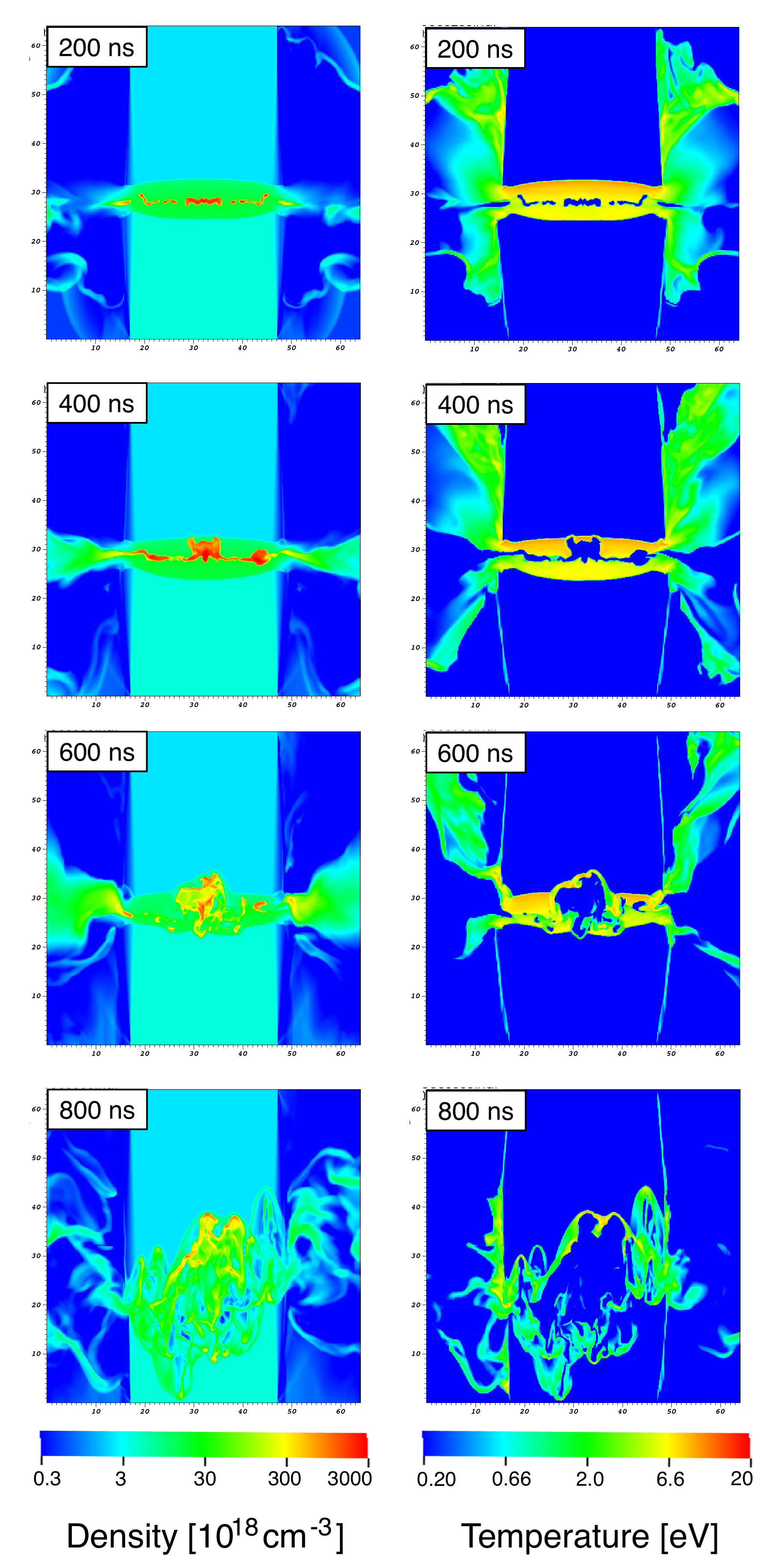}
    \caption{
    Run 9 from table \ref{tab:params}, at intervals of 200 ns. Density is plotted in the left column, while the right column shows a plot of temperature. At later times the Nonlinear Thin Shell Instability can be observed.
    }
    \label{fig:6}
\end{figure}

Lastly, we examine  effects of changing the jet velocity. While previous scenarios used a velocity of 70 km s$^{-1}$, the final set of runs uses velocities of 80 km s$^{-1}$ for the top jet and 60 km s$^{-1}$ for the bottom jet. Figure \ref{fig:5} shows three simulations with these velocities. The first case shows two jets of equal density; here the imbalance of velocity results in the interaction region moving with a velocity of \~ 10 km s$^{-1}$ after collision. 
The second case shows two jets of equal momentum density (and thus equal mass flux); the velocity of the interaction region is significantly reduced, but motion still occurs as a result of ram pressure imbalance. Both of these runs exhibit clumping behaviour similar to that of the runs with a density imbalance.
The final case shows two jets of equal ram pressure. The interaction region now remains stationary. While clumping behaviour is initially similar to the other two cases, clump mergers in this case produce a full cold slab spanning the entire radius of the jet. At later times (figure \ref{fig:6}), this cold slab is broken apart by the nonlinear thin shell instability, similar to the cases examined in \citet{Markwick21}.

\section{Discussion} \label{sec:discuss}

\subsection{Motion of the Interaction Region}\label{sec:velocity}

As seen in figures \ref{fig:2}, \ref{fig:3}, and \ref{fig:5}, the interaction region has a nonzero velocity unless the ram pressure $\rho v^2$ is equal for both the top and bottom jet. Otherwise, it moves away from the jet with higher ram pressure with velocity $V$. In a reference frame comoving with $V$, the momentum flux entering the interaction region from one jet must balance the momentum flux entering from the other jet. This constraint can be expressed as 
\begin{equation} \label{eq:p_flux_balance}
   \rho_1  \left( v_1 - V \right)^2 + p_1=  \rho_2  \left( v_2 - V \right)^2  + p_2
\end{equation}
where positive velocities indicate the downward direction; $v_2$ is therefore negative. We shall also neglect the thermal pressure difference $p_1-p_2$.

The requirement that both jets flow into the interaction region produces the constraint
$v_2 < V < v_1$. Taking the square root of equation \ref{eq:p_flux_balance}, subject to this constraint gives us
\begin{equation}\label{eq:ram_balance}
  \sqrt{\rho_1}\left( v_1 - V \right) =  \sqrt{\rho_2} \left(V - v_2 \right)
\end{equation}
which can be solved for $V$:
\begin{equation} \label{eq:v_interact}
V = \frac{ v_1 \sqrt{\rho_1}  + v_2 \sqrt{\rho_2} }{\sqrt{\rho_1}+\sqrt{\rho_2}}
\end{equation}

We will now examine this result in the context of figures \ref{fig:2}-\ref{fig:5}.
Velocities can be measured as follows: the location of the cold slab, lower jet shock, and upper jet shock are measured at $t=130$ ns and $t=230$ ns. The positions at each time step are averaged; taking the difference of these averages between the two time steps gives us the displacement (positive is down) over the 100 ns time interval.  For runs 7-9, a similar procedure is used except that the measurements are taken at $t=140$ ns and $t=280$ ns. Results are given in table \ref{tab:velocities}.

\begin{table}
    \centering
    \begin{tabular}{|c||c|c|c|c||c|c|}\hline
        Run & $n_1$ & $n_2$ & $v_1$&$v_2$ & $V$ (eq \ref{eq:v_interact}) & $V$ (measured)\\\hline\hline
         0 & 3.00&3.00 & 70 & 70  & 0 & 0\\\hline
         1 & 3.00&2.00 & 70 & 70 &  7.07 & 6 \\\hline
         2 & 3.00&1.00 & 70&70 &  18.8 & 17 \\\hline
         3 & 3.00&1.69 & 70&70 &  9.97 & 9\\\hline
         4 & 3.00&6.75 & 70&70 &  -14.0 & -12\\\hline       
         5 & 3.00&3.00 & 70&70 & 0 & 0\\\hline
         6 & 3.00&3.00 & 70&70 &  0 & 0\\\hline
         7 & 3.00&3.00 & 80&60 &  10.0 & 9\\\hline
         8 & 2.63&3.50 & 80&60 &  5.0 & 5 \\\hline
         9 & 2.30&4.08 & 80&60 &  0 & 0\\\hline
    \end{tabular}
    \caption{Densities and velocities of the jets in each run, along with the estimated (equation \ref{eq:v_interact}) and observed value for the interaction region velocity $V$. All velocities are given in km s$^{-1}$ and densities are given in $10^{18}$ cm$^{-3}$}
    \label{tab:velocities}
\end{table}

We can  connect this prediction to the experiments of  \citet{suzukiVidal15}, in which the jet velocities were measured to be $v_1 = 80\pm 10$ km s${ }^{-1}$ and $v_2 = -60\pm 10$ km s${ }^{-1}$.  Substantial uncertainties ($\sim 60\%$) were noted for measurements of density, but in any case the ratio $\frac{\rho_1}{\rho_2}$  was within an order of magnitude of unity. 
If equation \ref{eq:p_flux_balance} is instead solved for $\frac{\rho_1}{\rho_2}$ we find
\begin{equation}
\frac{\rho_1}{\rho_2} = \left(\frac{v_2-V}{v_1-V}\right)^2
\end{equation}
The bow shock velocity is stated to be $40\pm 10$ km s${ }^{-1}$, so the density ratio evaluates to $\frac{\rho_1}{\rho_2} = 6.25 \pm 4.38$; this on the higher end of what is in agreement with the measurements presented in \citet{suzukiVidal15}.

Another useful quantity to compute for collisions between flows of unequal ram pressure is the Mach number in the centre-of-mass frame. In this paper we define Mach number to be $M^2 = \frac{\rho v^2}{\gamma p}$, and $M_1$ and $M_2$ to be the Mach numbers for top and bottom flows in the laboratory frame. When combined with equation \ref{eq:v_interact} this definition gives us 
\begin{equation}
M^2 = \frac{\rho_1(v_1-V)^2}{\gamma p_1} = \frac{\rho_1}{\gamma p_1}\left(v_1-\frac{v_1\sqrt{\rho_1}+v_2\sqrt{\rho_2}}{\sqrt{\rho_1}+\sqrt{\rho_2}}\right)^2
\end{equation}
for the centre-of-mass frame (note that Mach numbers must be equal when ram pressures balance). This simplifies to
\begin{equation}
M^2 = \frac{\rho_1\rho_2}{\gamma p_i}\left(\frac{v_1-v_2}{\sqrt{\rho_1}+\sqrt{\rho_2}}\right)^2
\end{equation}
which, assuming $v_2 < 0 < v_1$, can be further simplified to 
\begin{equation} \label{eq:M_interact}
M = \frac{M_1\sqrt{\rho_2}+M_2\sqrt{\rho_1}}{\sqrt{\rho_2}+\sqrt{\rho_1}}
\end{equation} 
Note that while $v_2$ is assumed to be negative, we always define the Mach number to be positive. Note that while the interaction region velocity is an  average  weighed by the square roots of densities, the  Mach number is an average weighted by the square roots of \textit{inverse} densities.

\subsection{Cold Slab in One Dimension}

We now begin our examination of shocked lateral outflow by considering a scenario with no outflow: the growth of the cold slab for a flow colliding with a wall in one dimension. In subsequent sections, these results will be useful in the development of a three-dimensional model.

\begin{figure}
    	\includegraphics[width=\columnwidth]{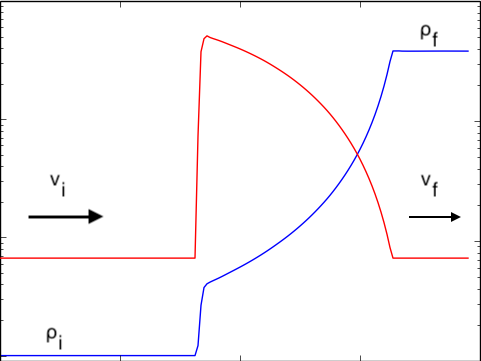}
    	\includegraphics[width=\columnwidth]{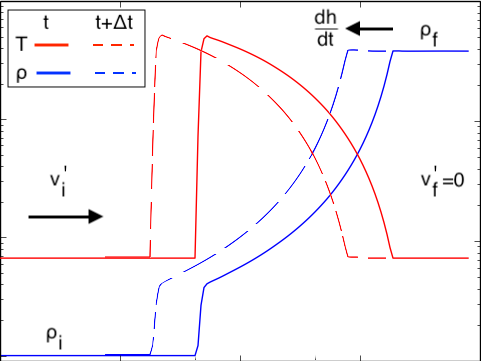}
    \caption{
    Temperature (red) and Density (blue) for a one dimensional shock in a) the shock frame and b) the lab frame.
    In the shock frame, the shock remains stationary while material in the cold slab flows away from the shock. In the lab frame, the shock moves to the left (dashed curve) while material already in the cold slab remains stationary.
    }
    \label{fig:slabFrames}
\end{figure}

Define $\rho_i, v_i$, and $T_i$ to be the initial pre-shock density, velocity, and temperature; also define $\rho_f, v_f,$ and $T_f$ to be the same values for material in the cold slab.  
Velocities are specified in a ``shock" frame (see figure \ref{fig:slabFrames}), in which the front (i.e. the side closer to the shock) of the cold slab remains stationary\footnote{ we assume that the radiative shock instability can be ignored, so that the distance between the shock and the front of the cold slab remains fixed}.
Any growth of the slab appears in this frame as the far side of the slab moving further away from the shock.

Let  $h$ be the height of the cold slab and $m = \rho_f hA$ be the mass of the cold slab within area $A$.
Mass continuity tells us that $\frac{\operatorname{d} m}{\operatorname{d} t}$ is equal to the incoming mass flux $\rho_f v_f$ times the area $A$, so
\begin{equation} \label{eq:mass1D}
\rho_f v_f A = \rho_f A \frac{\operatorname{d} h}{\operatorname{d} t} + Ah\frac{\operatorname{d} \rho_f}{\operatorname{d} t}
\end{equation}
which can be solved for $\frac{\operatorname{d} h}{\operatorname{d} t}$
\begin{equation} \label{eq:slabGrowth}
\frac{\operatorname{d} h}{\operatorname{d} t} =
v_f - \frac{h}{\rho_f}\frac{\operatorname{d} \rho_f}{\operatorname{d} t}   
\end{equation}. Going forward we shall assume that $\rho_f$ reaches a constant value once $h$ reaches an appreciable length,
allowing us to neglect the $\frac{h}{\rho_f}\frac{\operatorname{d} \rho_f}{\operatorname{d} t} $ term; we therefore conclude that the rate of growth of the slab is equal to the velocity of fluid flow within the slab. With no way for material to exit the slab, the slab will grow indefinitely. 

We now establish a relation between the preshock and cold slab states. Conservation of momentum requires that 
\begin{equation}
    \rho_i v_i^2 + p_i = \rho_f v_f^2 + p_f
\end{equation}
Using mass continuity $\rho_i v_i = \rho_f v_f$ gives
\begin{equation}
v_f\left(v_i^2 + \frac{p_i}{\rho_i}\right) = v_i\left(v_f^2 + \frac{p_f}{\rho_f}\right)
\end{equation}
Under the assumption of an isothermal shock ($\frac{p_f}{\rho_f} = \frac{p_i}{\rho_i}$), this has two solutions:
$v_f = v_i$ (preshock), and $v_f = \frac{p_i}{\rho_i v_i}$ (cold slab). For the latter solution mass continuity then gives  
\begin{equation}
    \rho_f = \frac{\rho_i v_i}{v_f} = \frac{\rho_i^2 v_i^2}{p_i}.
\end{equation}
Since the Mach number in the shock frame is defined as $M^2 = \frac{\rho v_i^2}{\gamma p_i}$, we can express the state of the cold slab as\footnote{Note that the pressure in the cold slab exceeds the immediate postshock pressure by a factor of $\frac{\gamma+1}{2}$.}
\begin{equation} \frac{\rho_f}{\rho_i} = \frac{p_f}{p_i} = \frac{v_i}{v_f} = \gamma M^2. \label{eq:slabState}\end{equation}

We can apply this result to estimate the density of the cold slab in our simulations; we will do this for run 0. The density of the incoming flow is $3\times10^{18}$ cm$^{-3}$ while the incoming pressure is $2.98\times 10^{5}$ erg cm$^{-3}$. Using a mass of 1 amu, we find a sound speed of 3.15 km s$^{-1}$, which for a jet velocity of 70  km s$^{-1}$ gives us a Mach number of 22. This gives us a cold slab density of $2.4\times10^{21}$ cm$^{-3}$, which is in good agreement with what is observed in our simulations.

Before moving on to three dimensions, we briefly discuss this problem in the ``lab" frame, which is useful to consider for a flow running into a stationary wall. If we return to figure \ref{fig:slabFrames}, we see that in this frame the far side of the cold slab remains stationary while the shock moves away from the wall. We therefore can define $v_i' = v_i - v_f$ and $M' = \frac{v_i'}{v_i}M$ as the preshock velocity and Mach number in this frame. Using this, we find that equation \ref{eq:slabState} becomes
\begin{equation} \frac{\rho_f}{\rho_i} = \frac{p_f}{p_i} = \frac{v_i'}{v_f}+1 = \gamma M'^2\left(1+\frac{v_f}{v_i'}\right)^2\end{equation}
the solution of which is given by
\begin{subequations}
\begin{equation} \frac{v_f}{v_i'}= \sqrt{\frac{1}{4}+\frac{1}{\gamma M'^2} }-\frac{1}{2}, \end{equation}
and \begin{equation} \frac{\rho_f}{\rho_i}= \gamma M'^2 \left[\frac{1}{2}+ \sqrt{\frac14+\frac{1}{\gamma M'^2} }\right] +1. \end{equation}
\end{subequations}
These expressions can also be derived
by considering conservation laws in the lab frame, noting that the mass of any region enclosing the shock increases over time.

\subsection{Cold Slab for Cylindrical flows}\label{sec:cylinder}

We now wish to extend the model developed in the previous section to a cylindrical model in which gas is permitted to escape laterally. Our primary aim in this section is to use mass conservation to estimate the size of the cold slab.  Although some gas also escapes from the sides of the cooling regions without ever reaching the cold slab, the cooling region  accounts for just 10\% of the mass of the interaction despite being around 90\% of the volume. We therefore make the approximation that the mass of the SLO originates entirely from the cold slab material.  A more accurate calculation could be done by dividing the cooling region into  layers and computing the SLO for each layer.  

Let $r_-$ be the radius of the incoming flow (the reason for labelling as minus will become apparent in the next section).
Assume $h$ is the height of our cylinder, and that mass flows into the cylinder from both the top and bottom and out the sides. The outgoing mass flux is equal to $\rho_h (2\pi r_0 h)v_\perp$, so equations \ref{eq:mass1D} and \ref{eq:slabGrowth} 
 become
 \begin{subequations}
  \begin{equation}  \label{eq:slabGrowth2a}
\rho_f v_f (2\pi r_-^2) -\rho_hv_\perp(2\pi r_- h) = \rho_f (\pi r_-^2) \frac{\operatorname{d} h}{\operatorname{d} t} + \pi r_-^2h\frac{\operatorname{d} \rho_f}{\operatorname{d} t}
\end{equation}
 \begin{equation} \label{eq:slabGrowth2}
\frac{\operatorname{d} h}{\operatorname{d} t} =
2v_f - \frac{h}{\rho_f}\frac{\operatorname{d} \rho_f}{\operatorname{d} t} - \frac{\rho_h (2\pi r_- h)}{\rho_f(\pi r_-^2)} v_\perp.
\end{equation}
 \end{subequations}
In equilibrium this gives  an expression for the SLO velocity 
 \begin{equation} \label{eq:id_mass}
v_\perp =\frac{\rho_fv_f r_-}{\rho_h h}. 
\end{equation}

\begin{figure}
	\includegraphics[width=\columnwidth]{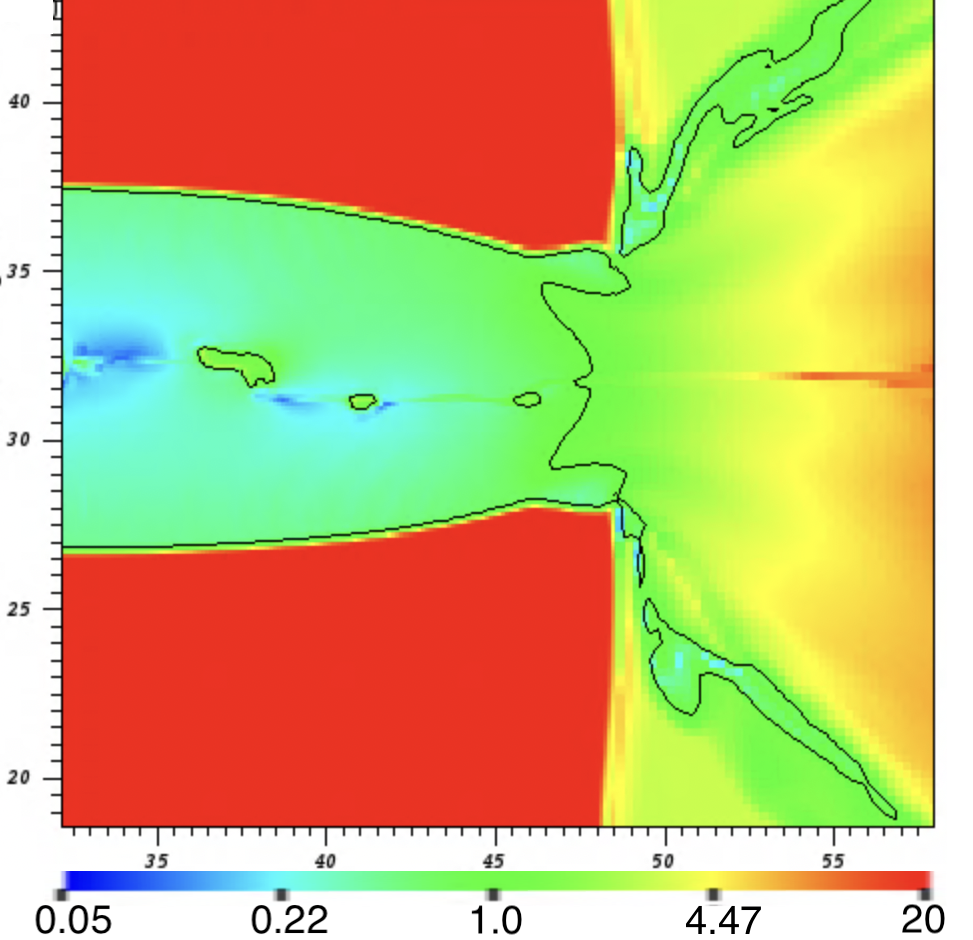}
    \caption{
    A plot of Mach number near the edge of the interaction region, for run 0 at time 660 ns. A black contour is drawn at $M=1$.
    }
    \label{fig:Mach}
\end{figure}

Let us begin by guessing that $\rho_h = \rho_f$. We will also {assume} that $v_\perp$ is equal to the sound speed (which must equal that of the incoming flows). This  is justifiable both theoretically \citep{Raga93} and by observation of our simulations (see figure \ref{fig:Mach}). We therefore find
 \begin{equation} 
 \frac{v_i}{M} = v_f \frac{r_-}{h}
\end{equation}
Since the slab has no net growth we can use equation \ref{eq:slabState} for $\frac{v_f}{v_i}$. Solving for $\frac{h}{r_-}$ therefore gives us
\begin{equation} \frac{h}{r_-} = \frac{1}{\gamma M}\label{eq:geom}\end{equation}
The actual ratio may be slightly smaller as a result of  aforementioned small amount of gas leaking laterally out of the cooling region. The jets in our {simulations}
have $\gamma = \frac53$ and $M\sim20$ so this factor should be approximately 0.03. For a jet of radius 1.5 mm the size of the cold slab would be just 0.045 mm, which is comparable to what we observe. We must note however that such a length corresponds to less than 4 cells on the grid at the highest level of refinement, so numerical errors are likely to be significant.

\subsection{Deflection Angle for Jets with Different Radii}\label{sec:angle}

Finally, we shall examine the deflection angle for SLOs produced by jets with different radii. We will assume that the jets are of equal density and velocity. Under these assumptions, the deflection angle can be approximated as the collision between the SLO of a system of identical jets with radius $r_-$ (the radius of the smaller jet), and an incoming flow from the larger jet between $r_-$ and $r_+$ (the radius of the larger jet). 

\begin{figure}
	\includegraphics[width=\columnwidth]{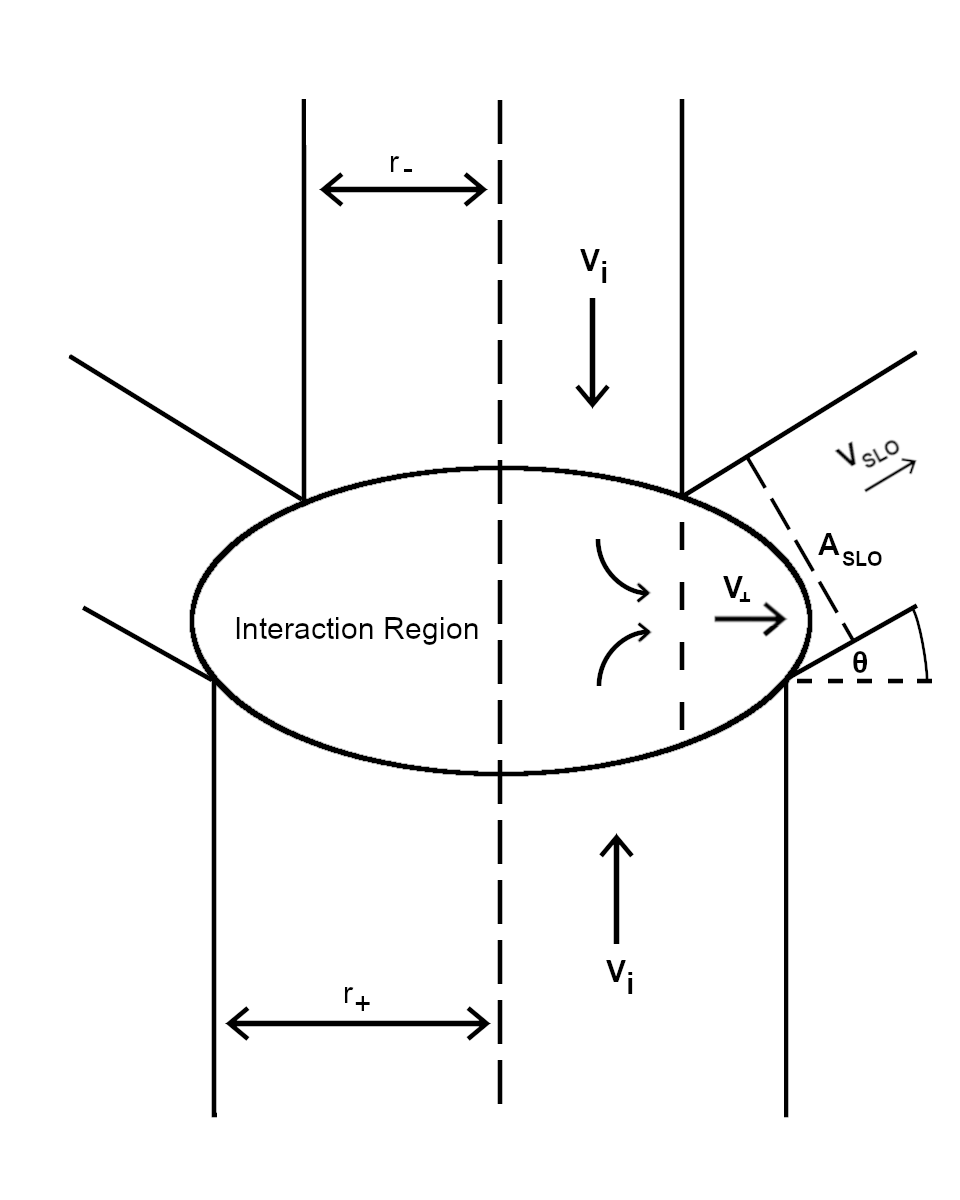}
    \caption{
    A diagram discussing the geometry of the problem for determining properties of the SLO for multiple jets.
    }
    \label{fig:diagram}
\end{figure}

Define angle $\theta$ to be the angle of deflection, relative to the surface normal to the incoming flows {of density $\rho_i$, speed, $v_i$, and pressure $p_i$} (see figure  \ref{fig:diagram}).
Define area $A_\text{SLO}\operatorname{d}\varphi$ as the area of some surface which is normal to the SLO and covers the extent of the SLO through angle $\operatorname{d}\varphi$ about the jet axis.
Let the SLO be characterized by $\rho_\text{SLO}$, $v_\text{SLO}$, and $p_\text{SLO}$; note that the $\text{SLO}$ subscript denotes material in the deflected SLO, while an $h$ subscript describes a horizontal SLO.
The {outgoing} horizontal and vertical momentum fluxes must therefore be \begin{subequations}\label{eq:fluxes}
 \begin{equation}
     F_x = (\rho_\text{SLO}v_\text{SLO}^2+p_\text{SLO})A_\text{SLO}\cos\theta \operatorname{d}\varphi
 \end{equation}
 \begin{equation}
     F_y = (\rho_\text{SLO}v_\text{SLO}^2+p_\text{SLO})A_\text{SLO}\sin\theta \operatorname{d}\varphi
 \end{equation}
\end{subequations}

We now express the horizontal {momentum flux 
leaving
 the interaction region 
 in terns if the incoming quantities}.
The SLO from the collision within the inner radius can be characterized as having density $\rho_h = \rho_f = \gamma M^2 \rho_i$  and velocity $v_\perp = \sqrt{\frac{\gamma p_i}{\rho_i}}$ (see equation \ref{eq:slabState} and the discussion in section \ref{sec:cylinder}).
The pressure $p_h$ can be estimated using conservation of energy within the cold slab (which features no cooling).
\begin{equation}
\left[\frac{\gamma p_f v_f}{\gamma -1}+\frac{\rho_fv_f^3}{2}\right](2\pi r_-^2) = \left[\frac{\gamma p_h v_\perp}{\gamma -1}+\frac{\rho_hv_\perp^3}{2}\right](2\pi r_-h)
\end{equation}
Note that we use $v_f$ (the state of material entering the cold slab) instead of $v_i$ as we are only considering energy flux in and out of the cold slab rather than the interaction.
Expressing $\rho_h$ and $v_\perp$ in terms of $\rho_f, p_f$, and $M$ allows us to solve for $p_h$
\begin{equation}
p_h = \left[\frac{3-\gamma}{2} +\frac{\gamma-1}{2\gamma^2M^2}\right] p_f
\end{equation}
Therefore in the limit of high Mach number, the horizontal momentum flux produced by the cold slab is equal to  \begin{equation} \label{eq:horiz_1}
F_x^h = (\rho_h v_\perp^2 + p_h) (r_- h \operatorname{d}\varphi) \approx Mp_i \left(\frac{3+\gamma}{2}\right) (r_-^2 \operatorname{d}\varphi)
\end{equation}

While neglecting the cooling region is sufficient when considering mass flux, the high pressures found in the cooling region also contribute significantly to the momentum flux. We can however continue to ignore the kinetic term $\rho_\text{cr} v_\perp^2$ as this term approaches a constant in the limit of high Mach number and thus accounts for less than one percent of the momentum flux for collisions with $M\sim 20$.
We shall approxmiate the pressure within the cooling region to be equal to the immediate postshock pressure; this pressure within the cooling region should exceed this, but by a factor of less than $\frac{\gamma+1}{2}$ (assuming pressure in the cooling region never exceeds that of the cold slab).
We therefore conclude that a lower bound for the contribution of the cooling region to the momentum flux is given by
\begin{equation} \label{eq:horiz_2}
F_x^L = p_\text{ps} (2r_- L \operatorname{d}\varphi) = \frac{4\gamma M^2 \xi}{\gamma+1} p_i(r_-^2 \operatorname{d}\varphi)
\end{equation} where $L$ is the (vertical) size of the cooling region and  $\xi = \frac{L}{r_-}$. 
In \citet{Markwick21} we showed how to calculate $L$ for a power-law cooling function as an improvement of the approximation given in \citet{CI}, and a similar approach can be used for a more complex function.  While performing such a calculation is necessary if one wishes to predict the outflow angle without running a simulation, when comparing analytical predictions to simulation results we will instead take the approach of measuring $L$ from the simulation results.

In the vertical direction, the incoming jet flux is
\begin{equation} \label{eq:vert}
F_y = (\rho_iv_i^2+p_i)\left(\frac{r_+^2 - r_-^2}{2} \operatorname{d}\varphi\right) = (\gamma M^2+1)p_i\left(\frac{r_+^2 - r_-^2}{2}\right)  \operatorname{d}\varphi
\end{equation}
Using equations \ref{eq:fluxes} we can deduce $\tan\theta = \frac{F_y}{F_x^h+F_x^L}$; combining this with equations \ref{eq:horiz_1} - \ref{eq:vert} allows us to estimate the deflection angle:
\begin{equation}\label{eq:angle}
\tan\theta = \left[\frac{(\gamma+1)(\gamma M^2 +1)}{8\gamma \xi M^2 + (\gamma+1)(\gamma+3)M}\right]  \frac{r_+^2 - r_-^2}{r_-^2}
\end{equation} 

We now compare equation \ref{eq:angle} to our simulations. 
Using $\gamma=\frac53$, $M=20$,  $L = 0.4$ mm,  $r_- = 1.0$ mm and $r_+ = 1.5$ mm, we find $\tan\theta = 0.93$ or $\theta = 43^\circ$; this is confirmed by the 600 ns frame of figure \ref{fig:4} which shows a deflection angle of $\theta = 41^\circ$. 
Meanwhile if we use $r_- = 1.5$ mm and $r_+ = 2.0$ mm we find $\tan\theta = 0.73$ or $\theta = 39^\circ$, with the measured angle being $36^\circ$.

\begin{figure}
    	\includegraphics[width=\columnwidth]{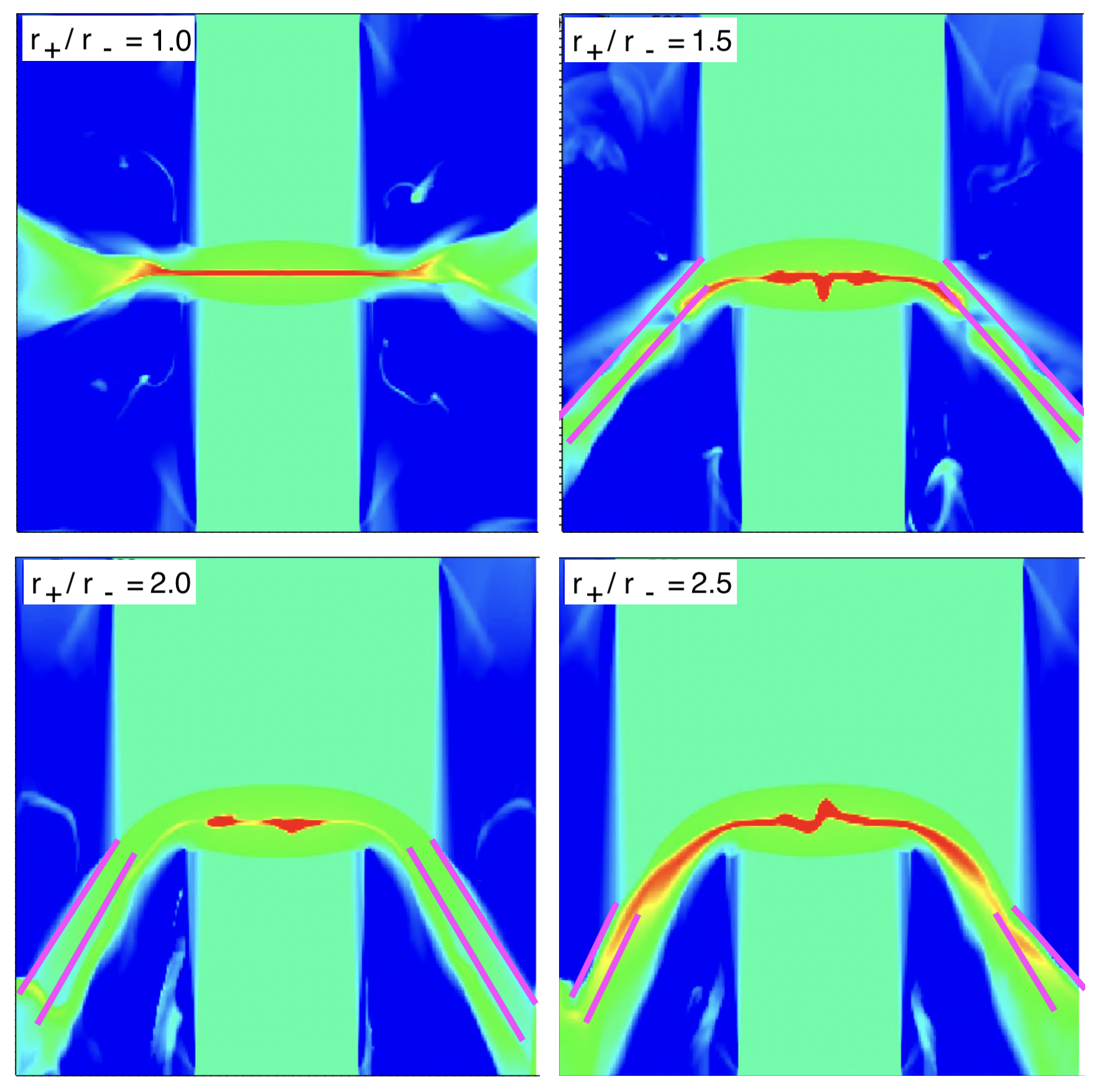}
    	\includegraphics[width=\columnwidth]{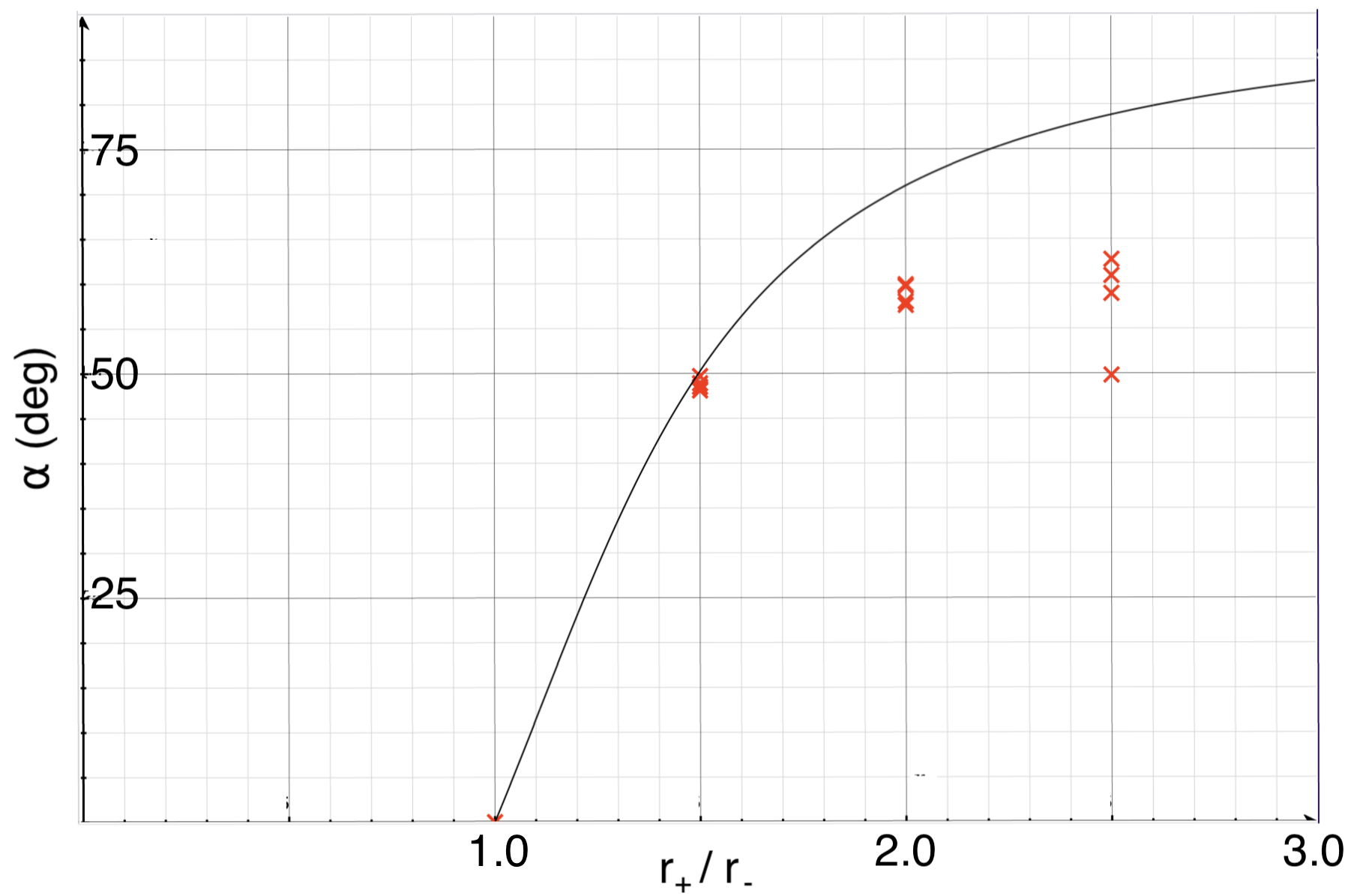}
    \caption{
    (top) A set of simulations used to further test equation \ref{eq:angle}.
    (bottom) Deflection angles vs the ratio of radii.
    Each red cross corresponds to a magenta line in the simulations, while the black curve is the estimated deflection. 
    }
    \label{fig:rad}
\end{figure}

To further test equation \ref{eq:angle}, we ran an additional set of simulations. These simulations were similar to run 6 (see figure \ref{fig:4}), but the radius of the top jet varied between values of 1.0 mm, 1.5 mm, 2.0 mm, and 2.5 mm; these runs were also conducted at lower resolution than the other runs. Using the 1.0 mm case we estimate $\xi = 0.3$, and for each subsequent run we took four measurements of deflection angle, shown in magenta in figure \ref{fig:rad}. We find that equation \ref{eq:angle} was very accurate for the 1.5 mm case, but sorely overestimates for $\frac{r_+}{r_-} \gtrsim 2$. An improved result may be obtained by considering the curvature in the region between $r_-$ and $r_+$, as equation \ref{eq:angle} is best suited to the limit where that region is small.

\subsection{Conclusions}\label{sec:conc}

In this paper we have continued our studies of colliding flows. While in \citet{Markwick21} we used identical flows and an analytic cooling model to isolate the effects of instabilities, in this paper we used a more realistic cooling function and variation of flow parameters; this was done in part to bring our simulations closer to the experiments of \citet{suzukiVidal15} while still remaining the hydrodynamic regime. Our simulations suggest explanations for two important features of the bow shock. First, for flows of differing density or velocity, the interaction region (and thus a shock which bounds it) moves at a speed given by equation \ref{eq:v_interact}. Second, we find that the conical shape of the bow shock arises from a mismatch in the cross-sectional radii of the flows, with the shocked lateral outflows being deflected away from the jet of larger radius.
All of the simulations in these first two papers have been conducted in the hydrodynamic limit. While this is useful for isolating various aspects of the underlying physics, magnetic fields are likely to be significant in both laboratory and astrophysical flows, and are thus a worthwhile inclusion in future papers on the subject. Effects relating to optical depth may also be worth exploring, particularly in connection to laboratory contexts.

\begin{acknowledgments}
This work used the computational and visualization resources in the Center for Integrated Research Computing (CIRC) at the University of Rochester. Financial support for this project was provided in part by the Department of Energy grants GR523126, DE-SC0001063, and DE-SC0020432, the National Science Foundation grant GR506177, and the Space Telescope Science Institute grant GR528562. Additional funding for this research was provided by the Center for Matter at Atomic Pressures (CMAP), a National Science Foundation (NSF) Physics Frontiers Center, under Award PHY-2020249. Any opinions, findings, conclusions or recommendations expressed in this material are those of the author(s) and do not necessarily reflect those of the National Science Foundation.
\end{acknowledgments}

\section*{Data Availability Statement}
The data that support the findings of this study are available from the corresponding author upon reasonable request.

\section*{Author Declarations}
The authors have no conflicts to disclose.


%
%

%


\bibliography{paper}

\end{document}